\documentclass[parskip=half]{scrartcl}

\usepackage{graphicx}

\usepackage[colorlinks=true]{hyperref}

\usepackage[sortcites]{biblatex}
\usepackage{bbding}
\usepackage{graphicx}
\usepackage{booktabs}
\usepackage{array}
\usepackage{amsmath}
\usepackage{orcidlink}

\begin{document}
    \title{How to design a Public Key Infrastructure for a Central Bank Digital Currency\thanks{This is an extended version of the paper accepted at SECRYPT 2025.}}
    \author{%
      Makan Rafiee%
      \thanks{secunet Security Networks AG, Kurfürstenstr. 58, 45138 Essen, Germany, \href{mailto:makan.rafiee@secunet.com}{makan.rafiee@secunet.com}}%
      \and%
      Lars Hupel%
      \thanks{Giesecke+Devrient GmbH, Prinzregentenstr. 161, 81677 München, Germany, \href{mailto:lars.hupel@gi-de.com}{lars.hupel@gi-de.com}}%
      \;\;\orcidlink{0000-0002-8442-856X}%
    }

\maketitle

\begin{abstract}
Central Bank Digital Currency (CBDC) is a new form of money, issued by a country's or region's central bank, that can be used for a variety of payment scenarios.
Depending on its concrete implementation, there are many participants in a production CBDC ecosystem, including the central bank, commercial banks, merchants, individuals, and wallet providers.
There is a need for robust and scalable Public Key Infrastructure (PKI) for CBDC to ensure the continued trust of all entities in the system.
This paper discusses the criteria that should flow into the design of a PKI and proposes a certificate hierarchy, together with a rollover concept ensuring continuous operation of the system.
We further consider several peculiarities, such as the circulation of offline-capable hardware wallets.
\end{abstract}

\section{Introduction}

\emph{Central Bank Digital Currency (CBDC)} is a digital means of payment, issued by a country's (or region's) central bank, denominated in the national currency. According to the latest results of the annual CBDC survey conducted by the Bank for International Settlements, 94\% of the respondents say they are working on digital currency \cite{DiIorio2024}. As of 2024, many major central banks are pushing forward with CBDC, including the European Central Bank with their \emph{Digital Euro} project.

While many of the projects are not yet in production stage, there is an emerging view that a full launch encompasses at least the following criteria \cite{Hupel2024}:

\begin{itemize}
    \item continuous and uninterrupted availability for an indefinite amount of time, i.e. no unannounced shutdown,

    \item real legal tender that can always be exchanged at face value with cash and deposit money,

    \item no system resets, i.e. holdings will remain valid,

    \item upgrade and maintenance work requires little to no intervention from users, except for long-term hardware upgrades, similar to the 2-5 year cycle of bank cards and smartphones.
\end{itemize}

\noindent
Because CBDC is public digital infrastructure, it needs to satisfy the highest resilience and security standards, at least on par with national settlement and payment systems.

But there are also additional requirements unique to CBDC: As opposed to traditional banking infrastructure, a CBDC would operate in 24/7 mode and has thus very little room for maintenance and/or downtime.

\paragraph{Structure of this paper}
The remainder of the introduction gives an overview over public key infrastructure (PKI) (§\ref{sec:intro-pki}) and explains why it matters for CBDC (§\ref{sec:intro-pki-matters}). Afterwards, we discuss related work (§\ref{sec:related-work}): while ample literature about both PKI and CBDC are available, little information is available how to design a PKI for a production-grade CBDC.

To set the stage for the rest of the paper, we further introduce the participants in a CBDC system (§\ref{sec:participants}). Based on this, we can design the PKI, taking into consideration pecularities of CBDC (§\ref{sec:pki}). Finally, we propose a certificate rollover concept (§\ref{sec:rollover}).

\subsection{Public key infrastructure}\label{sec:intro-pki}

The backbone of any digital currency is the correct use of cryptographic materials and choosing appropriate security levels. This includes selecting the right algorithms for encryption, key exchange, digital signatures, and hash functions. More succinctly: getting the cryptographic primitives right.

But just picking the right algorithm is not enough to build trust in the whole system. Key material needs to be uniquely and verifiably connected to all entities. Many entities need to be authenticated before payments can happen, for example:

\begin{itemize}
  \item wallets need to authenticate each other;

  \item the central bank want to ensure that only authorized wallets are used to hold currency; and

  \item commercial banks need to confirm that the counterparties they are exchanging money with are genuinely who they claim to be.
\end{itemize}

\noindent
\emph{Public Key Infrastructure (PKI)} is the solution to these problems. By mapping public key materials to abstract entities in a certificate, participants in the network can always rely on the authenticity of their communication partners. This requires a root of trust---a central authority---from which all certificates are derived. In the case of CBDC, this would be the central bank.

Trusting these derived certificates is a result of users trusting the central bank. For that, the central bank must have proper processes in place to ensure a secure certificate life cycle. This entails handling \emph{Certificate Signing Requests (CSR)}, rights and roles, and organisational procedures. RFC 3647 \cite{RFC3647} provides a good overview of key considerations for designing a PKI.

Focusing on CBDC, this paper explores PKI design decisions, authenticated entities requiring certificates, and how a seamless certificate rollover can be handled.

\subsection{Why is PKI significant for CBDC?}\label{sec:intro-pki-matters}

Outages caused by public key infrastructure, for example expired or wrong certificates, can cause significant damage. This affects many industry domains \cite{Hickman2024}.

For example, in 2023, a certificate with a mismatched top-level domain caused a short service interruption in Microsoft Sharepoint services \cite{Abrams2023}. While the problem was quickly fixed, it affected many customers across major services (including Microsoft Teams).

Hardware may also be impacted: in the same year, Cisco hardware was affected by certificates expiring after their 10 year lifetime \cite{Chervek2023}. Renewing certificates in hardware is notoriously complicated, since devices may not be directly connected to the internet and are therefore out of the manufacturer's control.

There is one known instance of a PKI issue impacting the operation of a CBDC. The Eastern Carribean Central Bank suffered from an outage in 2022, caused by expired certificates in \emph{Hyperledger Fabric}, the DLT employed by the bank \cite{Crothers2022}. As mentioned in the introduction, CBDC is a national payment infrastructure, therefore prolonged service interruptions can cause problems for the economy.

Assuming broad adoption of a CBDC in a given jurisdiction, and the potential use of digital currency in industrial applications, it becomes clear that a PKI supporting a currency's infrastructure must be highly resilient. Already today, many industries see ``an upward trend in certificate numbers'' \cite{Keyfactor2023}. CBDC wallets exacerbate the problem, because they rely on individual certificates per wallet.

\subsection{Contributions}

There are already numerous PKIs deployed and used.
Perhaps the most prominent example is the PKI used for internet, where multiple certificate authorities issue certificates in the X.509 format to a variety of servers, including for HTTP (web), SMTP (mail), and other protocols.

In this work, we argue that CBDC is an application that requires and deserves a dedicated PKI (§§\ref{sec:related-work}, \ref{sec:participants}).
The main reasons for that are:

\begin{itemize}
  \item existing PKI---particularly the decentralized internet PKI---is not specifically designed for the security needs of a CBDC system;
  \item the need for custom features and formats (see also §\ref{sec:pki:format});
  \item allowing the central bank to exert control over the system and its design;
  \item disaster recovery procedures.
\end{itemize}

\noindent
Further evidence to this special status is that today, many financial and/or government applications already use dedicated PKI.
Prime examples for this are the \emph{ICAO Public Key Directory} for validating passports and other digital identity documents \cite{SecureIdentityAlliance2022}, \emph{SwiftNet} for financial messaging, and \emph{SM-PKI} for smart electricity meters.

We contribute a comprehensive analysis of the security parameters of CBDC ecosystems, and from that, derive a PKI design that takes specifics of digital currencies into account.
The major difference to existing PKIs is the need to accommodate for unpowered, intermittently offline devices communicating directly with each other (§\ref{sec:rollover:specifics}), which greatly complicates the issuance and renewal of certificates in the field.
For example, current hardware-baed payment schemes like credit cards, do not offer person-to-person payments, and can therefore afford a simpler design.

Finally, we contribute a rollover concept for seamless operation of the system that could be generalized for other applications.

\section{Related work}\label{sec:related-work}

The literature has already acknowledged the need for robust and scalable Public Key Infrastructure for CBDC and other digital asset systems.

Several publications \cite{Chu2022,Christodorescu2020,Zhang2024,Yang2022} highlight the use of PKI for offline payments. The core idea is that wallets carry certificates signed by an authority, e.g. the central bank. These can be used for mutual authentication between a pair of wallets before or during a transaction. This establishes trust in an offline scenario. Illegitimate wallets are consequently excluded from the system and cannot inject counterfeit money. As Chu \emph{et al.} summarize, a PKI ``allows safe transactions even though the certificate authority is offline, since a certificate states whether the counterpart is a trusted user'' \cite{Chu2022}.

Takagari \emph{et al.} \cite{Takaragi2023} investigate the role PKI plays for \emph{Know-your-customer (KYC)} checks. The authors develop a privacy-enhanced PKI by which a ``financial institution verifies the identity of a prospective customer,'' based on the national ID, while simultaneously protecting the customer's personal information.

According to Han \emph{et al.} \cite{Han2019}, a PKI is a core ingredient for the regulatory layer of a CBDC. Their goal is ``the supervision of objects such as banks and third parties in network layer and users and transaction in user layer.'' Similarly, Zhang writes that the role of PKI is ``maintaining trust within the network of entities authorized to operate the CBDC system'' \cite{Zhang2024}.

Our previous research \cite{Hupel2024} identifies PKI as an instrumental part of any CBDC ecosystem. In particular, we have evaluated the various cryptographic keys and their algorithms involved, to understand how they are affected by the migration to post-quantum cryptography. This is echoed by Zhang, who explains that ``the PKI infrastructure adopted by the central bank should be quantum ready'' \cite{Zhang2024}.

In the related field of \emph{Distributed Ledger Technology (DLT)}, there is literature outlining the use of PKI for operating permissioned ledgers, i.e., where access control limits the participation in the system. For example, \emph{Hyperledger Fabric}, uses PKI for ``signature generation, verification, and authentication'' \cite{Campbell2019}. Pal \emph{et al.} \cite{Pal2021} survey some PKI strategies for major blockchain protocols.

The ISO has published \emph{Technical Report (TR)} 24374, entitled ``Security information for PKI in blockchain and DLT implementations''. It discusses ``the impact of different types of key management processes that are required for PKI implementations in Blockchain and DLT projects'' \cite{ISO24374}.

\section{Participants in a CBDC ecosystem}\label{sec:participants}

For this paper, we assume a CBDC system with the following characteristics:

\begin{itemize}
  \item \textbf{a two-tier distribution model}, i.e., the central bank manages the supply of money and commercial banks are responsible for distributing money to individuals and businesses;\footnote{%
    \emph{Project Aurum}\cite{BIH2022} distinguishes some further subtypes, but they are irrelevant for this paper%
  }
  
  \item \textbf{unbanked individuals are included}, i.e., people without a formal bank account can get access to CBDC, with notable groups being rural citizens without access to a bank branch, children, and tourists;

  \item \textbf{a wide spectrum of wallets are available}, allowing for both online and offline payments, especially hardware wallets embodied e.g. as smartcards;

  \item at least the payment scenarios \textbf{C2B/B2C} (customer-to-business/business-to-customer) and \textbf{P2P} (person-to-person) are supported;

  \item \textbf{a central bank ledger validates transactions}, with the concrete operational model (DLT/non-DLT) abstracted for this paper.
\end{itemize}

\begin{figure}[t]
  \includegraphics[width=\linewidth]{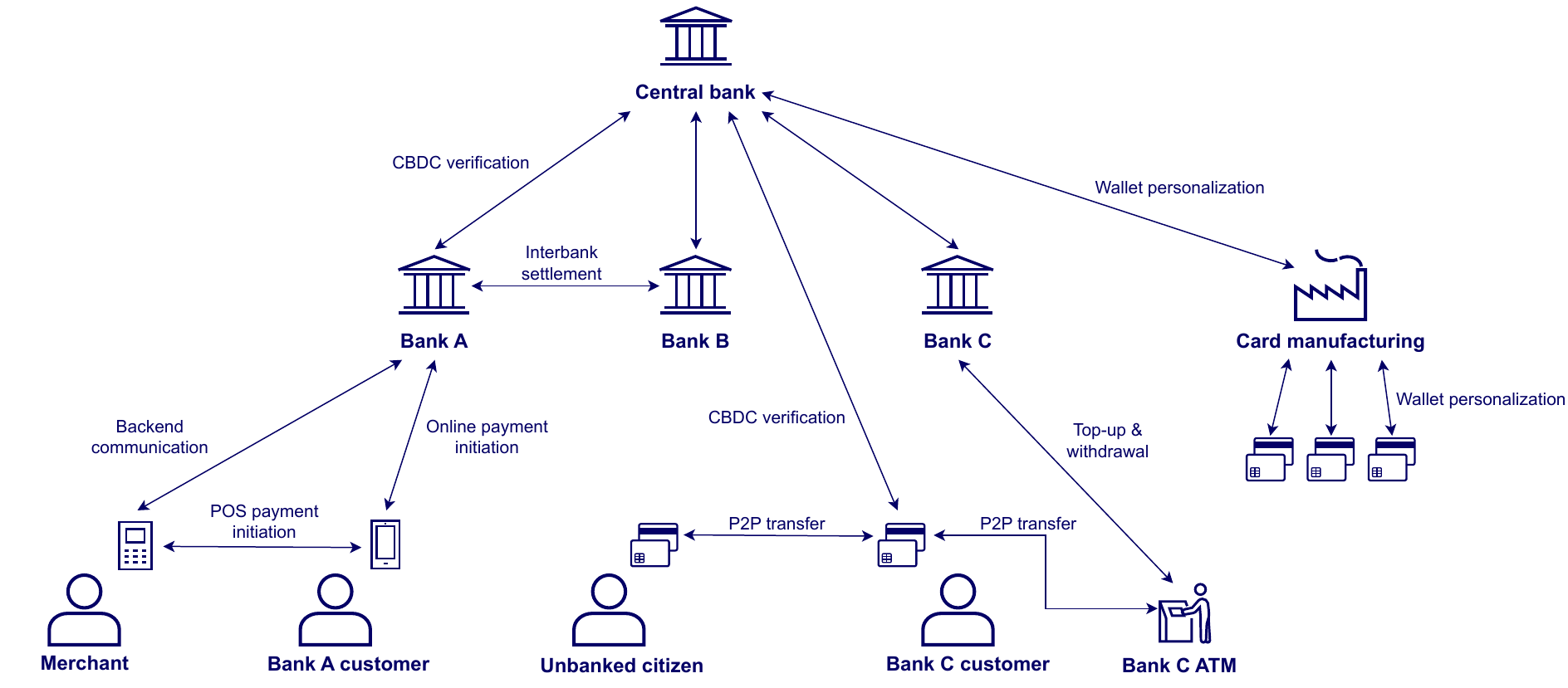}
  \caption{Participants and communication channels in a CBDC ecosystem}
  \label{fig:communication-channels}
\end{figure}

\noindent
Figure \ref{fig:communication-channels} shows the participants and communication channels.

The assumptions specified above give rise to a CBDC system that works similarly to cash. As the Hong Kong Monetary Authority points out, a CBDC is ``best designed as part of a two-tier system, with an appropriate division of labour between the central bank and private sector intermediaries for the distribution and circulation of CBDC'' \cite{HKMA2021}.

This section spells out the participants in the system and what their roles and responsibilities are: (1) the central bank, (2) financial service providers, and (3) individuals and businesses.

As noted above, wallets can come in different form factors, including hardware wallets such as smartcards.
They are typically manufactured by a dedicated entity (see also Figure~\ref{fig:communication-channels}).
But we set aside card manufacturing for now and will revisit this later in §\ref{sec:pki:smartcard}.

To keep it as simple as possible, we exclude a treatment of risk management, including fraud detection and investigation. This may require additional participants and communication channels.

\subsection{Central bank} 

The central bank focuses on ``providing the core, foundational infrastructure of a CBDC, guaranteeing the stability of its value and overseeing the system's security'' \cite{HKMA2021}. This includes, but is not limited to:

\begin{itemize}
  \item manage the monetary supply through minting and burning of CBDC;

  \item operate the central bank ledger and validating transactions \emph{(Central Register)};

  \item distribute CBDC to commercial banks;

  \item set rules and regulations for participants;

  \item operate the root certificate authority of the PKI.
\end{itemize}

\noindent
To a limited extent, the central bank could also provide CBDC wallets, for example to allow commercial banks the 1:1 exchange of central bank reserves for CBDC.

The central bank provides various communication channels.
Any participant can query the Central Register to validate CBDC.
It must also communicate with FSPs.

\subsection{Financial service providers}

Commercial banks, or more broadly, \emph{financial service providers (FSPs)}, are responsible ``to provide retail services to customers on a competitive basis, and leverage their network effects to innovate in business and service models'' \cite{HKMA2021}. Just like today, banks are the first point of contact for individuals and businesses. Their services include, but are not limited to:

\begin{itemize}
  \item perform user onboarding, e.g. by checking their identity, to comply with Know-your-customer (KYC) regulation;

  \item open and operate custodial customer wallets, and develop smartphone apps for customers to access their funds;

  \item issue hardware devices to customers and (potentially) unbanked individuals, to allow for offline payments and to enable unbanked people to participate in the CBDC ecosystem;

  \item distribute CBDC to their customers by providing exchange between deposit money, cash and CBDC;

  \item value-add services such as insurances.
\end{itemize}

\noindent
FSPs communicate directly with their customers, with the central bank, and with other FSPs.
They also provide communication channels to initiate and execute payments at the point of sale, as well as ATM operations, which may be proxied through existing systems and is therefore not specific to CBDC.

Note that some jurisdictions are considering the central bank to also act as a retail service provider, for example to aid financial inclusion. In other jurisdictions, non-banks (such as post offices) are thought to take over this role.

\subsection{Individuals and businesses}

Finally, the users of the system: individuals and businesses are exchanging CBDC with each other and/or use CBDC to pay for goods and services.

While details differ across jurisdictions, the most typical representatives of this group are local residents and merchants (both physical and online). Subject to various limits, they can obtain CBDC wallets and transact with each other. As opposed to traditional payment systems, CBDC payments are symmetric: a merchant's wallet will differ from an individual's wallet only in configuration.

Depending on the precise payment scenario, wallets will use different communication channels for executing payments.
In offline situations, it will typically be a connection based on proximity protocols such as NFC or Bluetooth, establishing direct peer-to-peer communication.
But there can also be online scenarios, where a user instructs their FSP to execute a payment.

\section{PKI concept}\label{sec:pki}

Just like cash, the trust in CBDC derives from the trust of participants in the central bank. This trust is the foundation of the secure operation of any CBDC ecosystem. As outlined above, such an ecosystem requires a number of entities.

These entities have a wide range of requirements. For example, some entities are operated by the central bank itself, while others are completely outside its control. This section explains the certificate hierarchy, as well as technical and procedural concerns.

\subsection{Hierarchy}

\begin{figure}[t]
  \includegraphics[width=\linewidth]{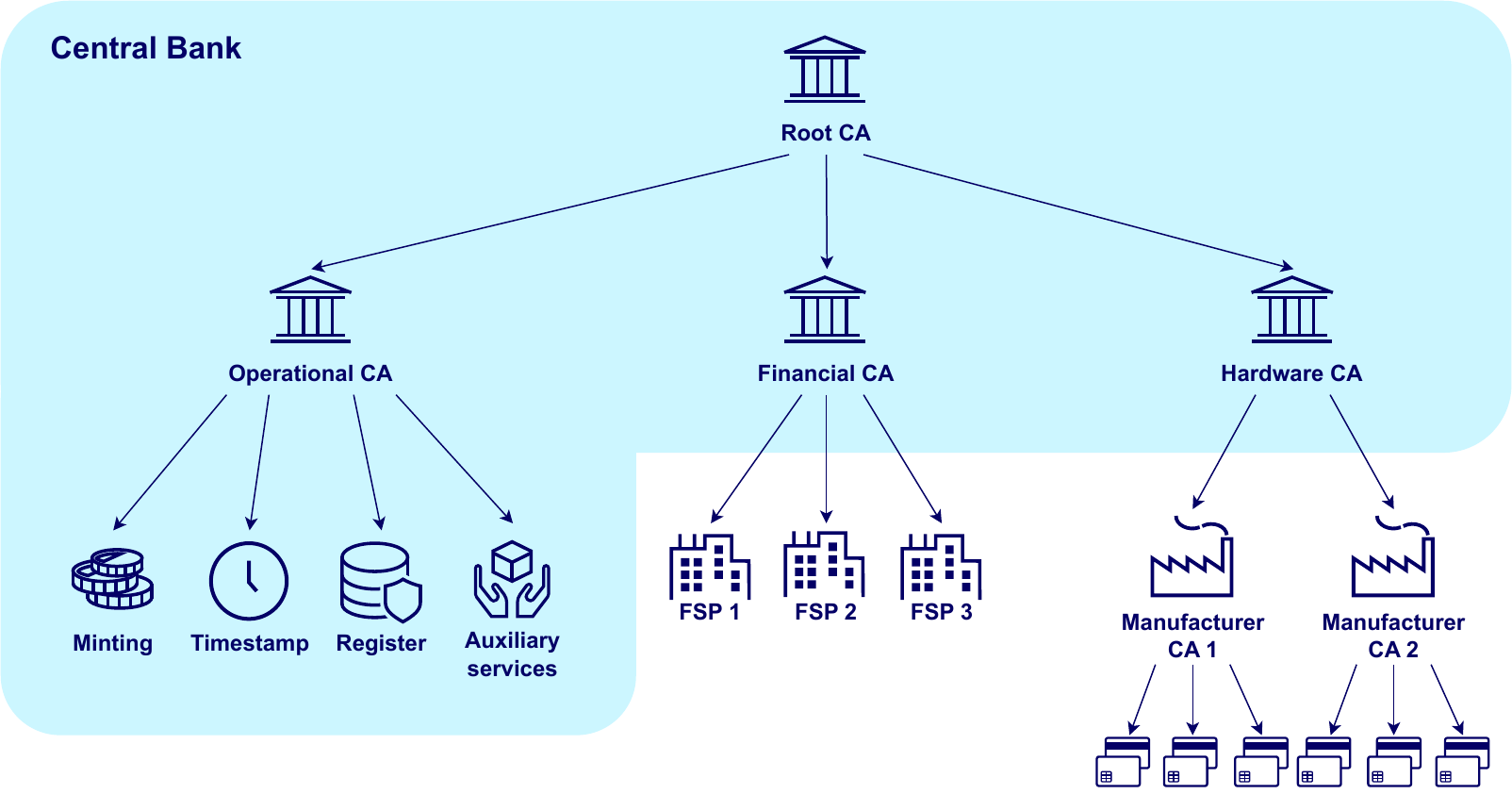}
  \caption{Proposed PKI hierarchy}
  \label{fig:pki-hierarchy}
\end{figure}

The central bank is the trust anchor of the whole system, but cannot perform all necessary processes required for a trustworthy system. It must, therefore, delegate trust towards other entities, such as financial service providers or manufacturers. That notwithstanding, the central bank still holds full responsibility for their correct operation, and therefore operates the \emph{Root CA}.

Therefore, the central bank imposes requirements to these thirds parties and enforces compliance. Ideally, the processes carried out by third parties shall be audited by the central bank.

The requirements in this policy could be published. This enables participants to understand how the PKI operates and establishes a factual basis for their trust.

Certificates with similar security levels should be clustered under the same \emph{Certificate Authority (CA)}, since certificates issued by the CA fulfill the same requirements towards issuance, revocation, delivery and further relevant processes.
We cluster certificates into three categories (Figure~\ref{fig:pki-hierarchy}).

\subsubsection{FSP certificates}

A financial service provider may only participate in the CBDC ecosystem after the central bank's approval. After onboarding, they would be equipped with a certificate that allows them to prove their identity to customers and other FSPs.

FSPs are authorized to open and operate wallets for customers. The FSPs will typically also have custody of private key material of their customers.
Whenever a payment across banks is made, the parties need to make sure that their respective counterparty holds a valid CBDC wallet.

From the user perspective, being a customer of an FSP already assumes (implicit) trust towards the FSP. Therefore, FSPs are already trusted to route payments to the desired receiver wallet, and no additional security gain can be achieved by equipping each online wallet with their own key material to authenticate the wallet. The necessary trust is already achieved by validating the FSP certificate, and let the FSP take care of routing and payment flows.

This way, there are trust relationships between a user and their bank, as well as between a bank and the central bank, but not between different banks.

\subsubsection{Smartcard certificates}\label{sec:pki:smartcard}

Similar to credit cards, the FSPs are responsible for distributing hardware wallets, such as smartcards, to their customers. 

In a CBDC ecosystem, peer-to-peer payments between wallets are allowed. This means that not all payments are routed through an intermediary. As a consequence, smartcards must be able to authenticate each other directly, which is especially important for offline payments.

The central bank bears ultimate responsibility that all smartcards are trustworthy. Since central banks typically do not manufacture those themselves, they will delegate this process to trusted smartcard manufacturers.

To guarantee flexibility and limit dependence on a singular smartcard manufacturer, multiple manufacturers should be supported.

Hence, the central bank should:

\begin{itemize}
  \item select a number of trustworthy smartcard manufacturers;

  \item issue and provide each smartcard manufacturer with an individual certificate;

  \item deliver the Root CA certificate securely to those manufacturers;

  \item ensure that the manufacturers comply with the rules and process requirements, which should happen contractually and with the help of a \emph{Certification Practice Statement} \cite{RFC3647};

  \item perform on-site audits on the facilities to ensure the manufacturers comply with the requirements (optional).
\end{itemize}

\noindent
Parts of this could also be delegated through FSPs, who may want to follow their own procurement procedures.

\subsubsection{Operational certificates}

The tasks that are performed by the central bank might be performed by different entities within the central bank itself. Naturally, this implies that multiple entities within the central bank need to hold individual certificates, for example:

\begin{description}
  \item[Minting] A system authorized to mint new money in a secure environment. There may also be personalized certificates for one or more central bank employees carrying private key material in a secure device.

  \item[Central Register] A ledger tracking authenticity of money, thereby confirming the validity of digital money; this system itself might be hierarchical (depending on choice of DLT or not).
  
  \item[Timestamp Service] A backend system that provides signed timestamps based on a trusted clock.
\end{description}

\noindent
The exact entities might differ depending on the broader design decisions for the CBDC.

\subsection{Certificate format}\label{sec:pki:format}

The state-of-the-art certificate format used in most industry use cases is \emph{X.509}~\cite{RFC5280}. In addition to subject identifier and public key, it provides many more certificate fields.

The X.509 standard has some drawbacks in environments with limited computation power, such as IoT devices or smartcards. Any auxiliary data and certificate fields slow down payments.

The obvious choice for a stripped down X.509 certificate format, specifically designed for smartcards, are \emph{Card Verifiable (CV)} certificates \cite{ISO7816-8}.
They enable smartcards to perform certificate operations with limited resources and power.
In contrast to X.509, CV certificates are much smaller and do not contain irrelevant information, thereby accelerating transaction processing.

\subsection{Revocation}

In any PKI, handling compromised certificates is crucial to ensure prompt invalidation and maintaining the trust and security of the system. There are two ways for this:

\begin{enumerate}
  \item Technical handling of compromised certificates (Revocation). The CA can revoke certificates in its system and provide this information to the relying parties\footnote{%
    \emph{Relying parties} are all entities validating certificates within a PKI. In a CBDC ecosystem, this can be many entities, including wallets.
  }
  in two ways:
    \begin{enumerate}
      \item \emph{Certificate Revocation Lists (CRLs)} are publicly available lists of revoked (but not expired) certificates. Whenever a certificate is validated, its revocation status can be looked up in the CRL. CRLs are issued (signed) by the responsible CA (or a dedicated CRL Signer).

      \item \emph{Online Certificate Status Protocol (OCSP)} allows the validating entity of the certificate to request the revocation status at a publicly available OCSP responder, which answers with good, revoked, or unknown.
    \end{enumerate}
  \item Organizational handling of compromised certificates. Certificates are not technically revoked, but the validating entities are manually informed about compromised certificates.
\end{enumerate}

\noindent
All of these approaches come with various problems.

CRLs have to be regularly updated, and revocation information may not be up-to-date at the time of the revocation check.

OCSP comes with a lot of additional traffic and, more importantly, with a significant privacy risk: OCSP requests are not encrypted and intercepting this traffic would reveal the certificates to be validated. This would clearly reveal communication metadata, weakening privacy.

Organizational handling of compromised certificates is difficult when the relying parties are numerous and unknown.

Payments demand high transaction throughput and low latency. Any additional data transfer within a transaction---including revocation status checks---must be carefully considered. Both CRL and OCSP lead to slower transactions.

Since the number of certificates in the PKI fairly limited, the organizational handling of compromised certificates should be considered instead. The relying parties are mostly controlled by the central bank, which means that in case of a malicious FSP, the central bank can simply unplug them from the system. The same is applicable to compromised operational certificates.

In offline payment scenarios, CRL and OCSP are not applicable for the lack of internet connection. OCSP stapling as an alternative method can be considered, even though they also increase the traffic and lead to longer transaction times.

\section{Certificate rollover}\label{sec:rollover}

In order to guarantee smooth transition between Root CA certificates---and therefore uninterrupted functionality of the PKI---rollover procedures are needed. For online entities, this is relatively easy, since certificates can be transmitted online for real-time updates. This is applicable to every entity in the PKI except for the smartcards.

Smartcards are offline entities which may, after production, rarely connect to an online service.  This gives rise to some additional complication that go beyond normal rollover procedures.

\subsection{Standard rollover}

\begin{figure}[t]
  \includegraphics[width=\linewidth]{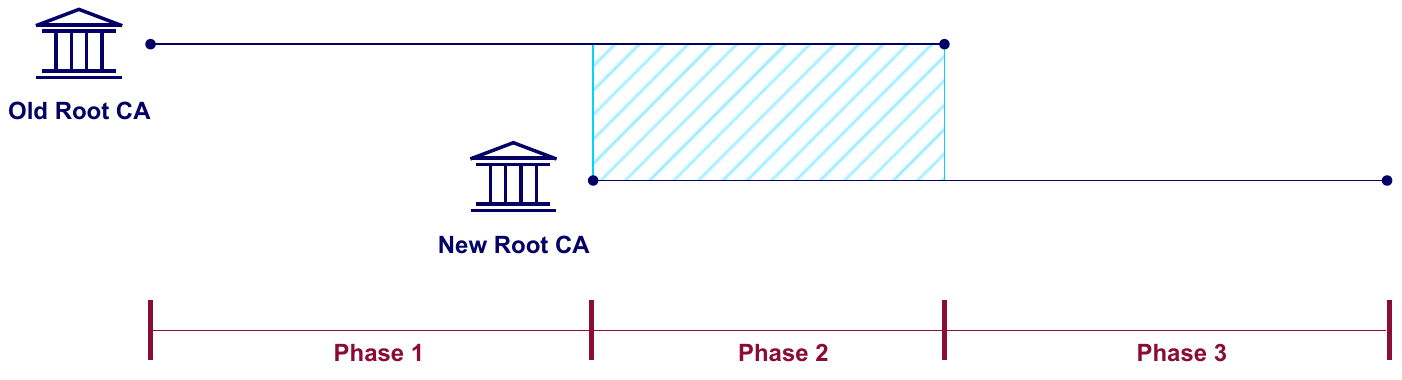}
  \caption{Conceptual CA rollover (standard case)}
  \label{fig:rollover:standard}
\end{figure}

Normally, a certificate rollover consists of three phases (see Figure~\ref{fig:rollover:standard}). In the first phase, only the old Root CA certificate is used to issue certificates and verify certificates based on the old Root CA certificate.

At the beginning of the second phase, a new Root CA certificate is created. New certificates in this period will be issued with the new Root CA certificate. Verification of certificates based of the old Root CA certificate will be verified with the old Root CA certificate, while certificates based on the new Root CA certificate will be verified with the new Root CA certificate.

Finally, only the new Root CA certificate is used to issue new certificates as well as verifying certificates. This completes the rollover. 

\subsection{CBDC and hardware wallet specifics}\label{sec:rollover:specifics}

Due to the nature of hardware wallets, certain complications need to be handled.

\subsubsection{Pinning}

In order to guarantee functionality and security over their lifetime, certificates should be pinned in smartcards. These pins could be especially protected, e.g. through hardcoding, to prevent tampering.

For the smartcards, the pinned certificates are:

\begin{itemize}
  \item Root CA certificate
  \item Timestamp Service certificate
  \item Central Register certificate
\end{itemize}

\noindent
In addition to the above-mentioned certificates, the smartcard holds its own certificate chain:

\begin{itemize}
  \item End-Entity (EE) (wallet) certificate
  \item Manufacturer CA certificate
  \item Hardware CA certificate
\end{itemize}

\subsubsection{Root CA ramp up}

The Root CA certificate is the trust anchor on which the PKI builds. When the Root CA certificate expires, every certificate based on that Root CA certificate loses its validity. For hardware wallets thay may be offline for extended amounts of time, a new Root CA certificate has to be created in advanced to it becoming active. Therefore, as compared to the standard rollover procedure (Figure~\ref{fig:rollover:standard}), we need to add an initial \emph{ramp up} phase: in this phase, the new Root CA certificate is issued, but not active (i.e., does not sign any certificates). Still, it would be installed alongside the old Root CA certificates in the smartcards.

\subsubsection{Smartcard manufacturer}

The certificate a manufacturer producer receives has a limited validity time. But, upon receiving a certificate, it should be able to manufacture smartcards over a period of timem independent of the issuing date. PKI rollover should be planed such that no change in the certificate chain should be needed after the initial phase, where all necessary certificates for production have been gathered. The validity of the smartcard manufacturer certificate has to be verifiable as long as there are smartcards in circulation based on that certificate.

\subsubsection{Smartcard validity period}

Smartcards that leave the manufacturer should always have the same lifetime, independent of their manufacturing date. Therefore, the complete certificate chain must be verifiable as long as those smartcards are circulating. This is true not only for online services, but also for mutual authentication.

\subsection{Solution requirements}

We define the validity period of a smartcard certificate, denoted as $1u$, as the base unit in our concept. This can be defined by the central bank and will typically be a few years. All other validity periods are derived from this unit.

Following the specifics for CBDC explained in the previous section, we consider the following requirements:

\begin{enumerate}
  \item All necessary certificates to conduct payments must be present on the smartcard for offline usage.

  \item Root CA, Central Register and Timestamp Service certificates must be pinned on the smartcard.

  \item A smartcard must be usable for its entire lifetime without updating pinned certificates.

  \item Smartcard manufacturer certificates must be able to be issued at any time.

  \item Smartcard manufacturers should be able to manufacture smartcards for a defined amount of time without updating their facilities with new certificates after receiving all necessary certificates from the central bank.
\end{enumerate}

\begin{figure}[t]
  \includegraphics[width=\linewidth]{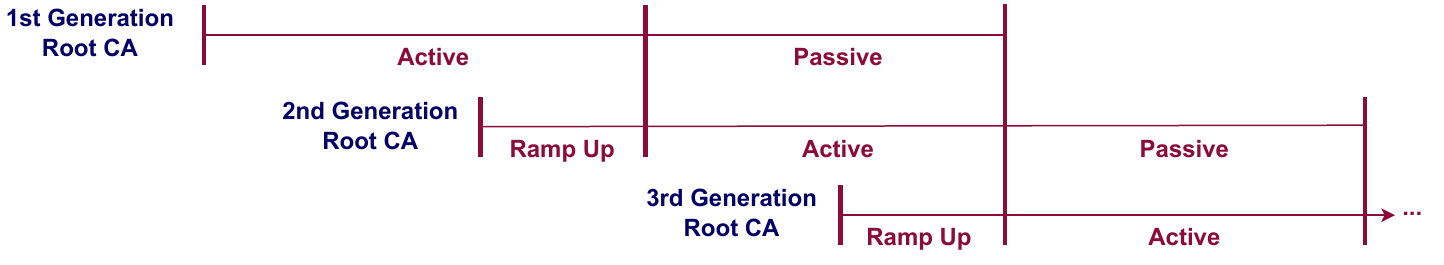}
  \caption{Conceptual CA rollover (CBDC case with ramp up phase)}
  \label{fig:rollover:cbdc}
\end{figure}

\noindent
This results in the necessity of a ramp up phase for certain certificates (Figure~\ref{fig:rollover:cbdc}.
The details are explained per certificate in the following sections.

\subsection{Hardware CA strand}

\paragraph{Manufacturer CA}

The smartcard manufacturer must be able to issue certificates at any point in time during its active phase.

The validity period of a the CA certificate is dependent on two factors. The active time is the period during which smartcards are manufactured based on a certain Root CA certificate. The passive time is the subsequent validity period of a smartcard, since functionality has to be guaranteed after manufacturing.

Therefore, the active and passive phases of the CA must each be at least as long as the smartcard validity period:
\begin{align*}
  \text{Active}(\text{Manufacturer CA}) &\geq u \\
  \text{Passive}(\text{Manufacturer CA}) &\geq u
\end{align*}

\paragraph{Hardware CA}

The same reasoning can be applied to Hardware CA certificates, since they also must be able to issue Manufacturer CA certificates at any point in time. Since their validity period is at least $2u$, with each of the phases being at least $u$, we get:
\begin{align*}
  \text{Active}(\text{Hardware CA}) &\geq 2u \\
  \text{Passive}(\text{Hardware CA}) &\geq 2u
\end{align*}

\subsection{Operational CA strand}

It must be ensured that all circulating smartcards pin the currently used Central Register certificate. Therefore, the \emph{ramp up} phase of the Central Register certificate must be at least the smartcard validity period.
\begin{align*}
  \text{RampUp}(\text{Central Register}) &= u \\
  \text{Active}(\text{Central Register}) &\geq u \\
  \text{Passive}(\text{Central Register}) &= 0 \\
\end{align*}
The same applies to the timestamp service:
It is highly recommended to use the same validity period.

\paragraph{Minting}

The minting certificate is not relevant for smartcards. The only requirement is that its validity never exceeds the validity of the Operational CA. A canonical validity period is therefore $2u$.

\subsection{Financial CA strand}

This strand is not critical, since its certificates are all issued online. We can therefore use $2u$ as the period for FSP certificates, and the same values for the Financial CA as the Operational CA.

\subsection{Root CA}

At the time of activation of the new Root CA certificate, all circulating smartcards must already know the new certificate. Therefore, the ramp up phase must be at least u.

Unlike the Hardware CA and the Manufacturer CA, the Root CA certificate does not need to be able to issue certificates at all times. Rather, it creates subordinate CA certificates only at specific points. The only requirement is that issued certificates are valid along their whole lifetime.

To ensure that smartcard certificates issued on the last day of validity of the Hardware CA, we obtain a passive phase for the Root CA of at least 2u.

To summarize:
\begin{align*}
  \text{RampUp}(\text{Root CA}) &= u \\
  \text{Passive}(\text{Root CA}) &\geq 2u \\
  \text{Active}(\text{Root CA}) + \text{Passive}(\text{Root CA}) &\geq 4u \\
\end{align*}
There is some degree of freedom in choosing the precise value of Active(Root). Figure~\ref{fig:rollover-example} gives a concrete example.

\subsection{Summary}

\begin{table}[t]
  \centering
  \begin{tabular}{p{0.25cm}p{0.25cm}p{0.25cm}p{2.5cm}>{\centering\arraybackslash}p{2cm}>{\centering\arraybackslash}p{2cm}>{\centering\arraybackslash}p{2cm}>{\centering\arraybackslash}p{2cm}}
    \toprule
    &&&&
    Ramp up &
    Active &
    Passive &
    Validity
    \\
    \midrule
    \multicolumn{4}{l}{Root CA} &
    $u$ &
    $\geq 2u$ &
    $\geq 2u$ &
    $\geq 5u$
    \\
    \cmidrule{2-8}
    & \multicolumn{3}{l}{Hardware CA} &
    &
    $\geq 2u$ &
    $\geq 2u$ &
    $\geq 4u$
    \\
    && \multicolumn{2}{l}{Manufacturer CA} &
    &
    $\geq u$ &
    $\geq u$ &
    $\geq 2u$
    \\
    &&& Smartcard &
    &
    $u$ &
    &
    $u$
    \\
    \cmidrule{2-8}
    & \multicolumn{3}{l}{Operational CA} &
    &
    $\geq 2u$ &
    $\geq u$ &
    $\geq 3u$
    \\
    && \multicolumn{2}{l}{Central Register} &
    $u$ &
    $\geq u$ &
    &
    $\geq 2u$
    \\
    && \multicolumn{2}{l}{Timestamp Service} &
    $u$ &
    $\geq u$ &
    &
    $\geq 2u$
    \\
    && \multicolumn{2}{l}{Minting} &
    &
    $2u$ &
    &
    $2u$
    \\
    \cmidrule{2-8}
    & \multicolumn{3}{l}{Financial CA} &
    &
    $\geq 2u$ &
    $\geq u$ &
    $\geq 3u$
    \\
    && \multicolumn{2}{l}{FSP} &
    &
    $2u$ &
    &
    $2u$
    \\
    \bottomrule
  \end{tabular}

  \vspace{1em}
  \caption{Rollover timing for each certificate. Some certificates have a mandatory ramp up phase. The total validity (last column) is the sum of the ramp up, active and passive phases.}
  \label{tab:timing}
\end{table}

Table~\ref{tab:timing} lists the minimum periods derived from the considerations in the previous sections.
For simplicity, all direct subordinate CAs (Hardware, Operational, Financial CAs) should have the same total validity period.
If possible, it also makes sense to set use the same active and passive periods, even if conceptually not strictly necessary.

\begin{figure}[p]
  \includegraphics[width=\linewidth]{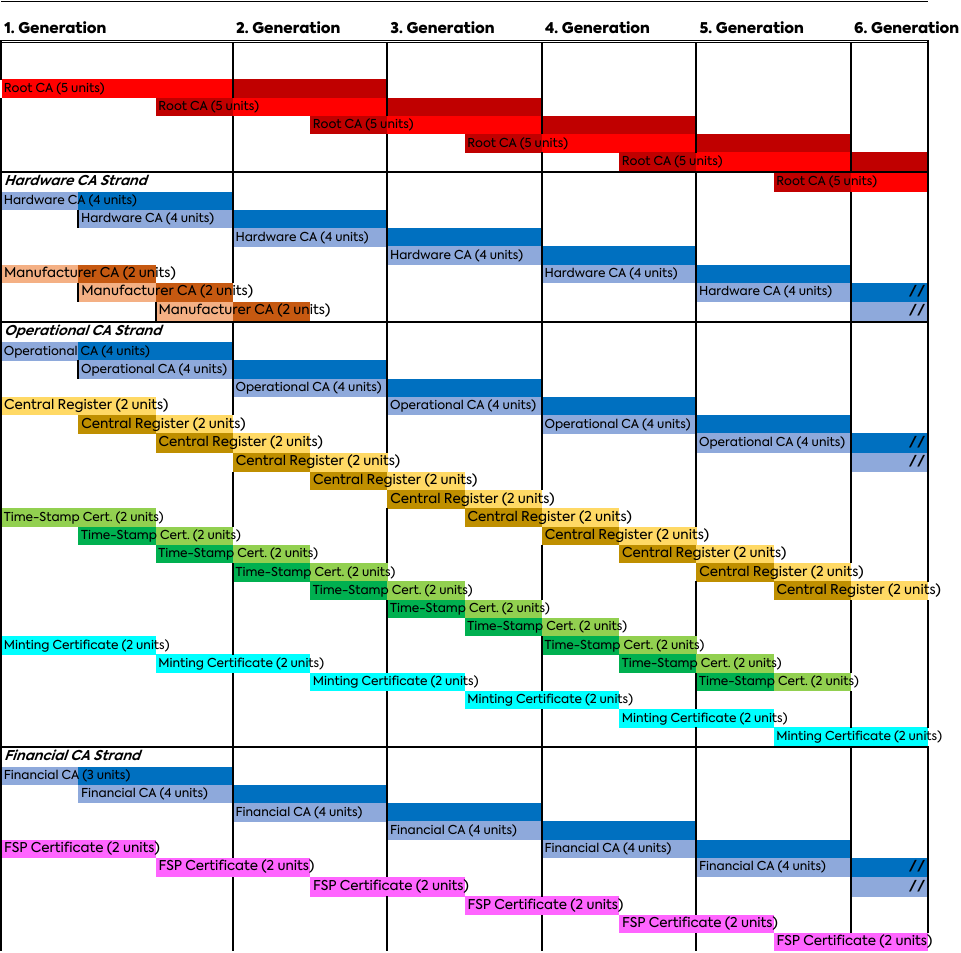}
  \caption{Example of a concrete rollover instantiation. Each certificate has a (light) active phase, and optionally a preceding (dark) ramp up phase and a succeeding (dark) passive phase.}
  \label{fig:rollover-example}
\end{figure}

Furthermore, Figure~\ref{fig:rollover-example} shows a concrete instantiation of the rollover concept.
The Root CA active phase is $3u$ in the first generation, and $2u$ in the subsequent generations.
(Subsequent generations also have a $u$ ramp up phase.)
All other certificate validity periods follow accordingly.

\section{Conclusion}

We have discussed how any CBDC system needs to authenticate entities and thus equip them with certificates. Since the trust in the system hinges on the trust in the central bank, we proposed to employ a Public Key Infrastructure (PKI) operated by the Central Bank. 

In particular, we introduced categories of certificates with similar security requirements and security processes, which lead us to a PKI design with concrete certificate authorities.

Furthermore, we investigated certificate formats and revocation procedures and concluded that stripped down X.509 certificates, such as CV Certificates, are appropriate for CBDC use. 

Technical certificate revocation procedures may negatively affect transaction throughput and performance. As an alternative, operational revocation procedures turned out to be most appropriate, since, in contrast to typical PKIs, relying parties are known beforehand. This leads to manageable revocation procedures at the relying parties, instead of a singe point of revocation status checks, such as OCSP responders and CRLs.

Finally, we proposed a rollover concept for a PKI, which enables seamless operation despite regularly scheduled certificate expirations. This rollover concept is generic and can be tailored to specific central bank needs. The smallest unit of validity is the smartcard validity period, which can be chosen to be any timeframe desired, and can be multiplied appropriately.

\paragraph{Acknowledgements} 
We thank our colleagues Nils Abeling, Klaus Alfert, Tolga Hazerli, and Peter Zeller for discussions on an early PKI concept.
This work has been partially supported by the Federal Ministry of Education and Research (BMBF), Verbundprojekt CONTAIN (13N16582).

\clearpage

\printbibliography

\end{document}